# Visualizing the Charge Density Wave Transition in 2H-NbSe$_2$ in Real Space

S.P. Chockalingam[1,*], C.J. Arguello[1,*], E.P. Rosenthal[1], L. Zhao[1], C. Gutiérrez[1], J.H. Kang[1], W.C. Chung[1], R.M. Fernandes[1], S. Jia[2], A.J. Millis[1], R.J. Cava[2], A.N. Pasupathy[1]

[1]Department of Physics, Columbia University, New York NY 10027
[2]Department of Chemistry, Princeton University, Princeton NJ 08540
* These authors contributed equally to this work

We report the direct observation in real space of the charge density wave (CDW) phase transition in pristine 2H-NbSe$_2$ using atomic-resolution scanning tunneling microscopy (STM). We find that static CDW order is established in nanoscale regions in the vicinity of defects at temperatures that are several times the bulk transition temperature $T_{CDW}$. On lowering the temperature, the correlation length of these patches increases steadily until CDW order is established in all of space, demonstrating the crucial role played by defects in the physics of the transition region. The nanoscale CDW order has an energy and temperature-independent wavelength. Spectroscopic imaging measurements of the real-space phase of the CDW indicate that an energy gap in NbSe$_2$ occurs at 0.7eV below the Fermi energy in the CDW phase, suggesting that strong electron-lattice interactions and not Fermi surface physics is the dominant cause for CDW formation in NbSe$_2$.

Transition-metal dichalcogenides are quasi-two dimensional materials, of which several show strong charge order at low temperature [1]. Our current understanding of the CDW transition from the normal state is based on momentum-space probes such as neutron [2] and x-ray scattering [3], angle-resolved photoemission spectroscopy (ARPES) [4, 5], optical spectroscopy [6], and Raman spectroscopy [7], which reveal a rather textbook-like second-order transition. Several pioneering STM spectroscopy studies have been performed deep in the charge ordered phase near zero temperature in both pristine and doped dichalcogenides [1]. However, none of the modern STM spectroscopic imaging techniques have been used to obtain a real-space, energy-dependent picture of either the CDW state or the CDW phase transition in the dichalcogenides. This requires precise, variable-temperature STM and scanning tunneling spectroscopy (STS) measurements through the phase transition [8]. Such studies on other complex materials like the cuprates and heavy fermions have yielded a wealth of information on the local electronic structure and have sparked many debates on the role of electronic inhomogeneity and charge order in these materials [9-11]. The dichalcogenides offer a clean test-bed to study the atomic-scale onset of a well-established CDW phase from the normal metal phase, without the complications of competing phases or strong disorder (intrinsic dilute disorder in the system will be shown to play a fundamental role on the nature of the CDW phase transition). We have performed such measurements on the prototypical CDW material NbSe$_2$, which has a bulk second-order phase transition from a normal metal phase to a nearly commensurate (3 X 3) charge ordered phase at $T_{CDW}$=33.7°K [2].

A typical STM image recorded at a temperature of T=38°K (Fig. 1(a)) shows three features that are common to all the images that we obtain. First, we see the surface selenium atoms which form a triangular lattice with lattice spacing of 3.44 Å. Also visible are surface and subsurface defects such as vacancies and interstitials. The density of such defects is in the range of 0.01-0.1% which is typical of the best crystals available. Finally, a short-range CDW is observed with a 3X3 atom periodicity in the immediate vicinity of the defects. At this temperature (T=1.2T$_{CDW}$) we see that the CDW covers approximately 50% of the surface area of the sample. Below T$_{CDW}$ (Fig. 1(b)), the CDW phase covers the entire sample area. Fig. 1(c)-(e) are STM images taken in the vicinity of single Se vacancies at temperatures of 57, 82 and 96°K. From these it is apparent that the short-range CDW can be observed in STM up to temperatures $\sim 3T_{CDW}$ .

Our STM measurements of this short-range CDW phase in NbSe$_2$ can be compared with transport measurements [12-15] on samples from the same batch. Shown in Fig. 1(f) is the temperature dependence of the in-plane resistance of one of such crystals. Consistent with previous measurements [12, 14], the resistance only shows a weak signature of the CDW transition, which is better visualized in the temperature derivative of the resistance shown in Fig. 1(g). A small additional contribution to the resistance appears below a temperature T$_O$ ~65°K; this is apparent in Fig. 1(g) as a decrease from the constant high T value of the derivative dR/dT (dashed line, Fig. 1(g)) when T<T$_o$. This additional contribution continues to steadily increase down to T$_{CDW}$, at which point the resistivity starts to drop quickly. These observations are completely consistent with our STM measurements. The presence of a short-range CDW phase would be expected to increase the resistivity by providing additional potential scattering to the normal state electrons. The strength of the short range CDW grows with decreasing temperature resulting in stronger scattering and increased resistivity, manifested by the reduction of dR/dT shown in Fig. 1(g). At T$_{CDW}$, the CDW establishes long-range order and thus the potential scattering from it is suppressed. Our STM measurements of the short-range CDW above the transition temperature thus point to the key role played by even weak disorder in determining the transport properties of complex materials.

The temperature dependence of the strength of the CDW order observed can be visualized by taking the Fourier transform (FT) of large-area STM images such as the one shown in Fig 1(a). Shown in Fig. 2(a) are two-dimensional FT images at 22, 38, 72 and 96°K after symmetrization along the high-symmetry directions. All of the images show hexagonal spots at the atomic peak position (red) and at the CDW ordering vector (indicated by arrows). The overall intensity of each image is normalized to equalize the atomic peak intensity at each temperature. Line cuts of the FTs along the atomic ordering vector are shown in Fig. 2(b). From this image, we see that the intensity of the CDW peak decreases with increasing temperature and becomes negligible around 100°K. The CDW wavelength is temperature independent within the resolution of the STM.

The correlation length for the short-range CDW also decreases with increasing temperature. In order to calculate the correlation length of the CDW at various temperatures, we Fourier filter the STM images to remove the atomic Bragg peaks, as described in the supplementary information. We then calculate the autocorrelation $A(x) = \frac{\int f(x')f(x-x')d^2x'}{\int f(x')f(-x')d^2x'}$ for each image. Fig. 2(e)-(f) are autocorrelation images of the CDW at various temperatures, with line cuts along the CDW modulation direction. We see that the correlation length decreases as the temperature increases. The extracted correlation length as a function of temperature is shown in the supplementary information.

A natural question that arises is whether the short-range patterns observed in the STM images are the precursors of a true broken symmetry or have a more mundane explanation. For example, short-range patterns are observed in the vicinity of point defects in a metal which arise from purely electronic standing waves due to quasiparticle interference (QPI)[16, 17]. To address the possibility that the observed patterns arise from QPI, we obtained STS maps over a wide range of electron energies at temperatures above $T_{CDW}$. A subset of one such data set is shown in Fig. 3(a)-(d), obtained at T=57°K. It is evident from the data that the periodic patterns around each defect are observed up to the highest energies measured (±1.4 eV), though the intensity and phase of the patterns is energy dependent. A 2D-FT of one of these images is shown in Fig. 3(e). This image displays sharp peaks at the atomic wavevector ($Q_{Bragg=}4\pi/\sqrt{3}a_0$) as well as at the CDW wavevector ($Q_{CDW} \sim Q_{Bragg}/3$). Similar FT images are obtained at all energies. Shown in Fig. 3(f) are line cuts at different energies of the 2D-FTs of STS maps along the atomic peak direction. We see from the line cuts that the CDW patterns seen in the STS maps have an energy independent wavelength.

We can compare our STS observations with what we would expect in the case of QPI. In general, QPI produces patterns in Fourier space that have intensity proportional to the joint density of states (JDOS)[17]. In order to model the expected patterns in Fourier space, we have performed density-functional theory (DFT) calculations [18, 19] of the band structure of NbSe$_2$ (calculation details are described in the supplementary information). Our results are in accord with other similar calculations [18, 20] as well as with recent ARPES measurements of the band structure [4, 5, 20, 21]. Shown in Fig. 3(g)-(j) are JDOS maps obtained at various energies from the calculations. It is evident that all of the structure seen in the JDOS maps is not seen in the STS maps; conversely, the sharp peaks seen in the STS maps at the CDW wavevectors are not seen in the JDOS maps. In simple terms, the band structure of NbSe$_2$ shows significant dispersion over the volt scale, and one would expect to see the effect of this dispersion in QPI measurements (QPI generically leads to energy dependent features). This is in direct contrast to what we observe in the experiment, indicating that the patterns seen in the experiment do not come from QPI. The fact that the observed CDW patterns show strong temperature dependence and persist out to Volt-scale energies indicates that a static lattice distortion is present in the short-range CDW regime and affects electronic states at all energies.

While the CDW wavelength remains constant over a wide energy range, the intensity and phase of the modulations seen in real space vary considerably for differing energies. We can directly see this from the STS maps shown in Fig. 3(a)-(d). In one dimension, the local phase of the CDW corresponds to whether a given point in real space corresponds to the crest (phase 0°) or trough (phase 180°) of the CDW modulation. The two dimensional CDW in NbSe$_2$ is therefore described by two local phases. Close observation of the images in Fig. 3(a)-(d) shows that the crests and troughs of the CDW do not align in space for all energies. In order to track the energy-dependence of the CDW phase, we calculate the two-dimensional cross-correlation between STS images at different energies and a reference image (for which we have chosen the highest energy STS map at E=+1.4 eV). We are thus able to follow the relative CDW phase with respect to this reference image. To visualize only the phases associated with the CDW, we band-pass filter the STS images around the CDW wavevectors to remove the effect of any other electronic inhomogeneity that may exist. A subset of these cross-correlation functions is shown in Fig. 4(a) (T=57°K). The cross-correlation functions at each energy can be fitted to extract the CDW phases, one of which is plotted in Fig 4(b) (the other phase has similar behavior). The CDW remains in phase from +1.4 eV down to about -0.7 eV, when it undergoes a ￼° phase shift, with a second weak reversal occurring beyond -1 eV. We eliminate the STM constant-current normalization as cause of the phase shifts by taking topographic images at different bias setpoints (see the Fig. S4 in the supplementary information) which between -0.6V and +1.0V look nearly identical whereas significant changes in the CDW are seen for images obtained with setpoints below -0.6V. These differences prove that real variations occur in the LDOS at large negative energies in a manner consistent with the spectroscopic data shown in Fig. 4(a)-(b).

To understand the implication of the change in phase of the CDW as a function of energy we analyze a minimal one dimensional tight-binding model of a CDW phase (see supplementary information for calculation details). Our model consists of two types of atoms (for example, Nb and Se to mimic $NbSe_2$), alternated in real space, with a single orbital on each atom. We assume that a CDW exists with a periodicity of 2 Se atoms, resulting in a 4 atom unit cell (2 Nb and 2 Se). Hopping is assumed to exist between neighboring Se and Nb atoms, as well as between adjacent Nb and Se atoms. We allow for both site centered order (different on-site energies of the two Se atoms) and bond centered order (asymmetric hopping parameters). For a given realization of the CDW, we calculate the contrast in the local density of states (LDOS) on the two adjacent Se atoms, since these are the surface atoms into which we tunnel. The sign of this quantity is equivalent to the phase measured in the STS experiments. Our calculations show that the phase changes take place at the energies corresponding to the largest CDW-induced mixing between states, i.e. at the energies where the gaps open, and the magnitude of the contrast is proportional to the size of the CDW gap. Our STS measurements therefore imply that strong spectroscopic changes happen at the CDW wavevector at an energy of -0.7 eV, with only weak changes at the Fermi energy.

In order to understand why the energy region around -0.7 eV is special in $NbSe_2$, we looked closely at the electronic structure of the material. Shown in Fig. 4(c) are the DFT bands projected on the Se states (green) and Nb states (red) at various energies along the CDW wavevector direction. These figures show that there is no significant nesting at energies near or above the Fermi energy (see supplementary information for plots at additional energies). However, at energies around -0.6 eV, we see that the electronic structure is strongly nested with the CDW wavevector (see Fig. 4 (d)). Thus at this energy range, we expect significant hybridization to occur in the CDW state along with the consequent opening of an energy gap. This key observation is in complete agreement with our experimental measurement of a change in the phase of the CDW in this energy range.

In the classic Peierls transition, the driving force for CDW formation is the electronic kinetic energy gain that occurs when strong Fermi surface nesting is present. In this mechanism, the important changes to electronic structure occur close to the Fermi energy (within a few $kT_{CDW}$), and higher energy states are inconsequential. Our experimental measurements indicate that the primary changes to the electronic structure when we enter the CDW phase are at much higher energies. This indicates that the Peierls mechanism is not the driving force for the CDW observed in $NbSe_2$. Rather, a strong coupling between the electrons and the lattice is essential for the formation of the CDW [22]. Our STS measurements for $NbSe_2$ also bear striking resemblance to the static patterns observed in the pseudogap phase of the cuprates [9-11, 23] that have been associated with charge density waves. However, a direct comparison between the two materials reveals two important differences. First, the cuprates show strong spatial patterns at energies within the pseudogap, with a diminishing strength at higher energies [10, 11]. Second, no real-space phase change occurs in the cuprate spatial patterns while crossing the cuprate pseudogap. These two facts together imply that the spatial pattern seen by STM in the cuprate pseudogap state is not a simple charge density wave gap of the type seen in $NbSe_2$.

Our findings also shed new light on the fundamental problem of second order phase transitions in the presence of randomness. The classical Wilson-Fisher paradigm [24] focuses on universal properties associated with diverging length and/or time correlations. Our new results point to aspects of the physics which, while non-universal in the Wilson-Fisher sense, are of great practical importance in understanding the actual phase transition behavior of dichalcogenides and other low-dimensional materials. Most importantly, we demonstrate that (as has been speculated theoretically [25]) the homogeneous normal state of this system is extremely fragile around impurities, leading to large patches of static order even very far from the phase boundary, despite the fact that impurities are very diluted in this system. This in turn fundamentally affects all macroscopic observables including transport coefficients [13, 14], scattering measurements [3], and spectroscopic properties [5]. Anomalous experimental measurements in these materials should be reinterpreted in light of our findings, and new theoretical tools should be developed for these materials in the regime of localized, static order.

**Acknowledgments**
We thank D. Efetov for experimental help and V. Oganesyan for discussions. This work has been supported by the US National Science Foundation under the Partnership for International Research and Education (RMF, AJM and ANP, grant number OISE-0968226) and CAREER (ANP, grant number DMR-1056527) programs and by DMR-1006282 (AJM), and by the Air Force Office of Scientific Research under grant FA9550-11-1-0010 (ANP). The crystal growth at Princeton was supported by the Department of Energy Basic Energy Sciences grant DE-FG02-98ER45706 (RJC).

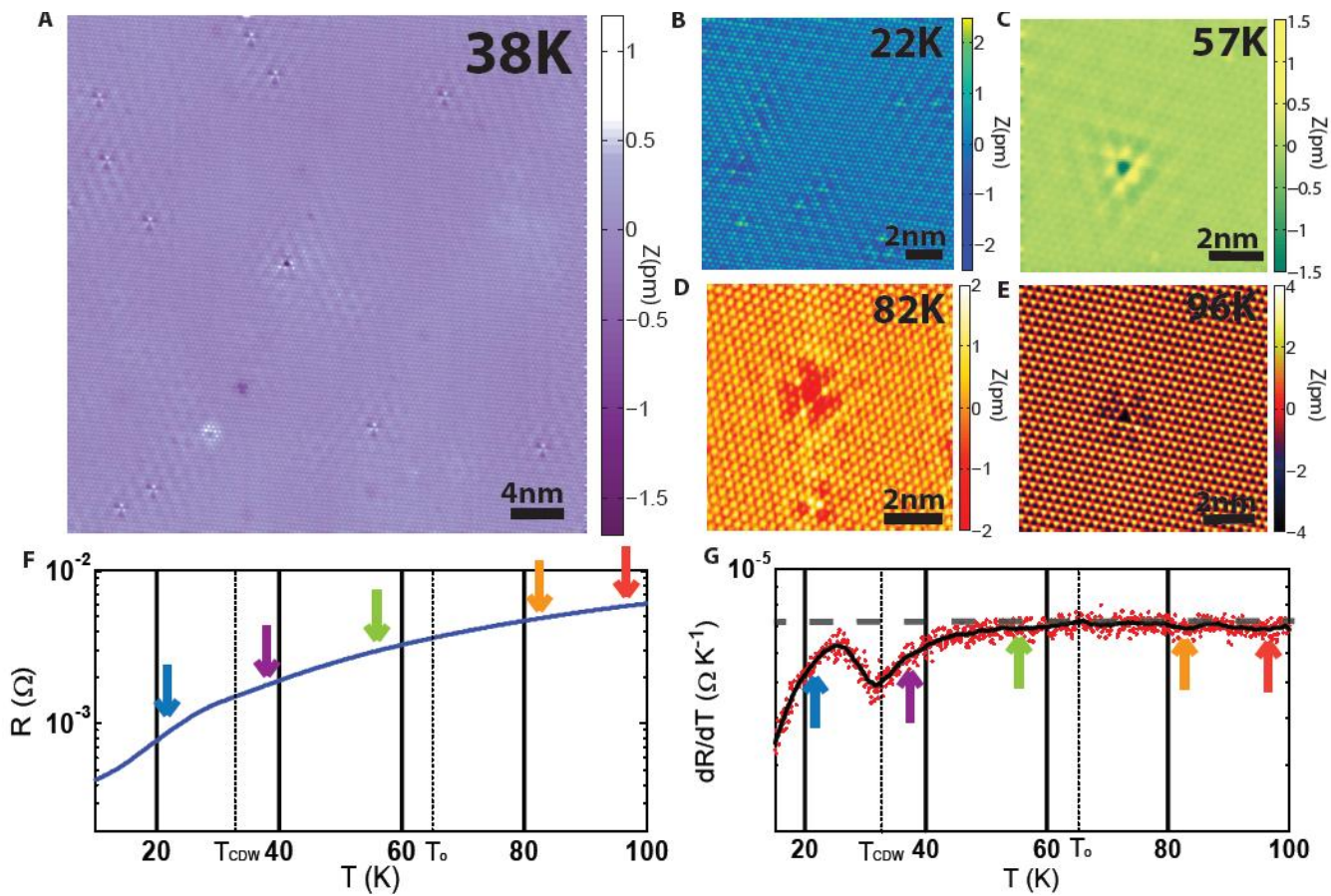

Fig. 1 (a) STM image (V=-100 mV, I= 20 pA) of the NbSe$_2$ surface above T$_{CDW}$=33.7°K showing surface Se, crystalline , and a short-range atomic superstructure (CDW). (b)-(e) STM images at various temperatures below and above T$_{CDW}$. (f) In-plane resistance and (g) temperature derivative of resistance of bulk samples plotted as a function of temperature. The temperatures at which the STM images are obtained are shown with color-coded arrows. On lowering the temperature below T$_O$ ~ 65°K, an additional contribution is seen in the resistivity (resulting in a decrease in dR/dT from the constant value at higher temperatures)

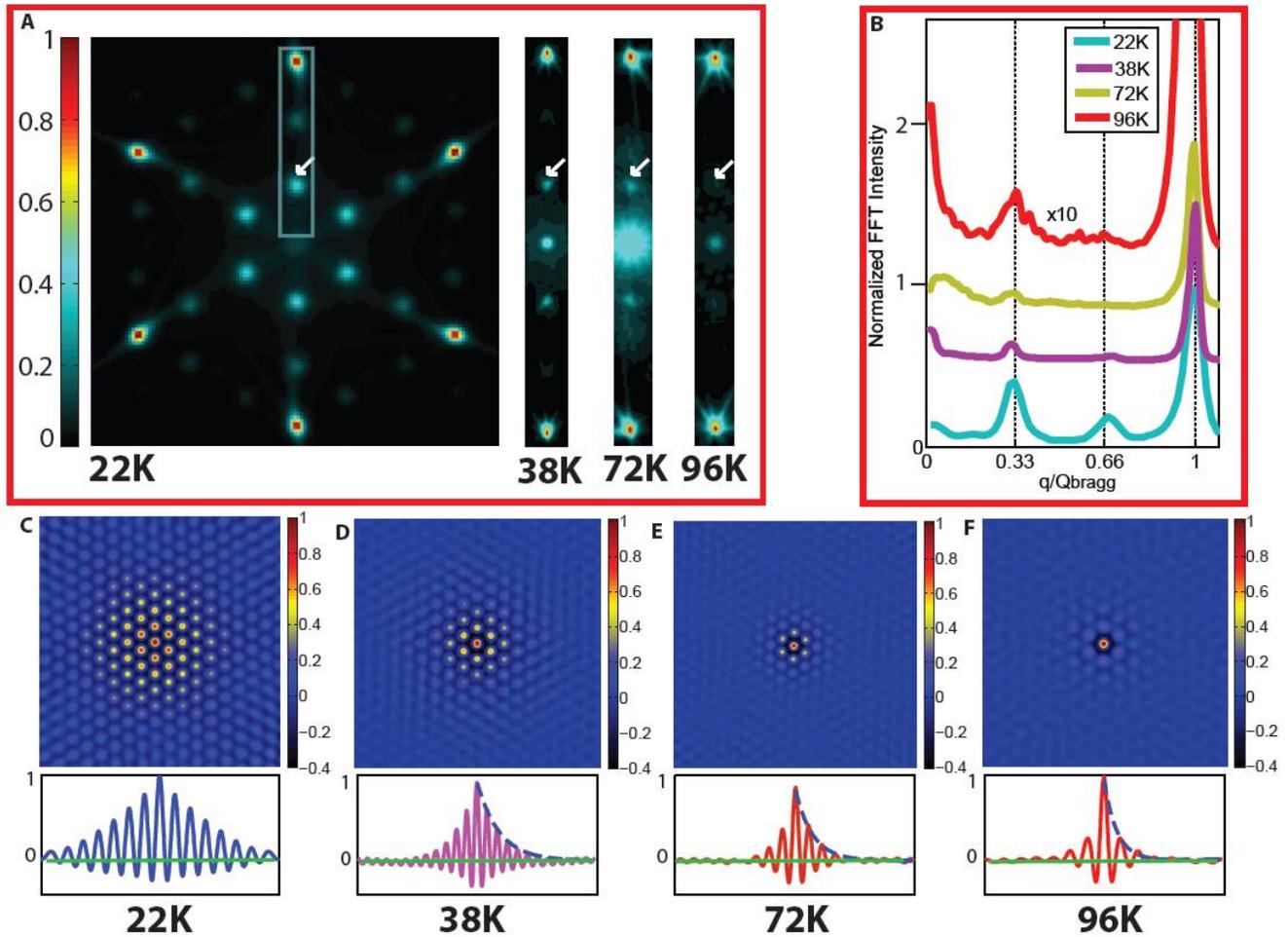

Fig. 2 (a) (left) FT of a 15 nm STM image obtained at 22°K. The red peaks correspond to the atomic peaks, and the CDW wavevector is marked by a white arrow. (right) Subsets of FT images of large-area STM maps at higher temperatures. The FT images are normalized to keep the atomic peak intensity constant. (b) Line cuts of FT images along an atomic wavevector at different temperatures (curves offset for clarity) showing the drop-off in CDW peak intensity with increasing temperature. (c)-(f) Autocorrelation functions and line cuts along the atomic directions from STM images at different temperatures. The images have been Fourier-filtered to remove the atomic peak. A correlation length can be extracted from the fall-off of the CDW intensity as a function of distance (dashed lines) at each temperature.

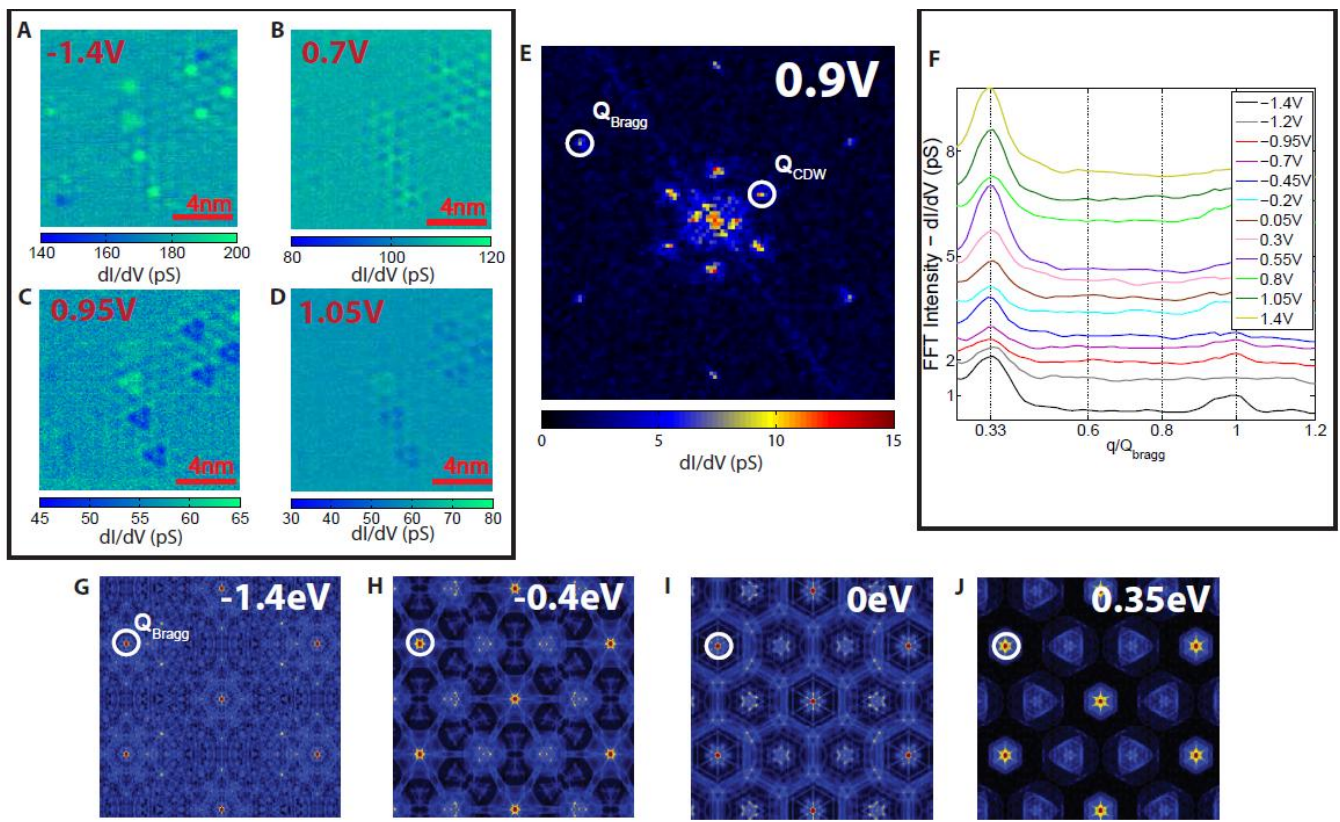

Fig. 3 (a) dI/dV maps taken at different energies at T=57°K over the same area of the NbSe$_2$ surface (V= -1.4 V, I= 700 pA). Short-range CDW around the atomic defects are present as well as local effects of the defects themselves. (e) FT image of the dI/dV map obtained at V=900 mV showing the atomic and CDW peaks but no other strong features at other wavelengths. (f) Line cut of FT images at different energies along the atomic peak direction. The CDW wavevector is seen to be independent of energy. (g)-(j) Expected scattering pattern (JDOS) from the calculated DOS of NbSe$_2$ at different energies.

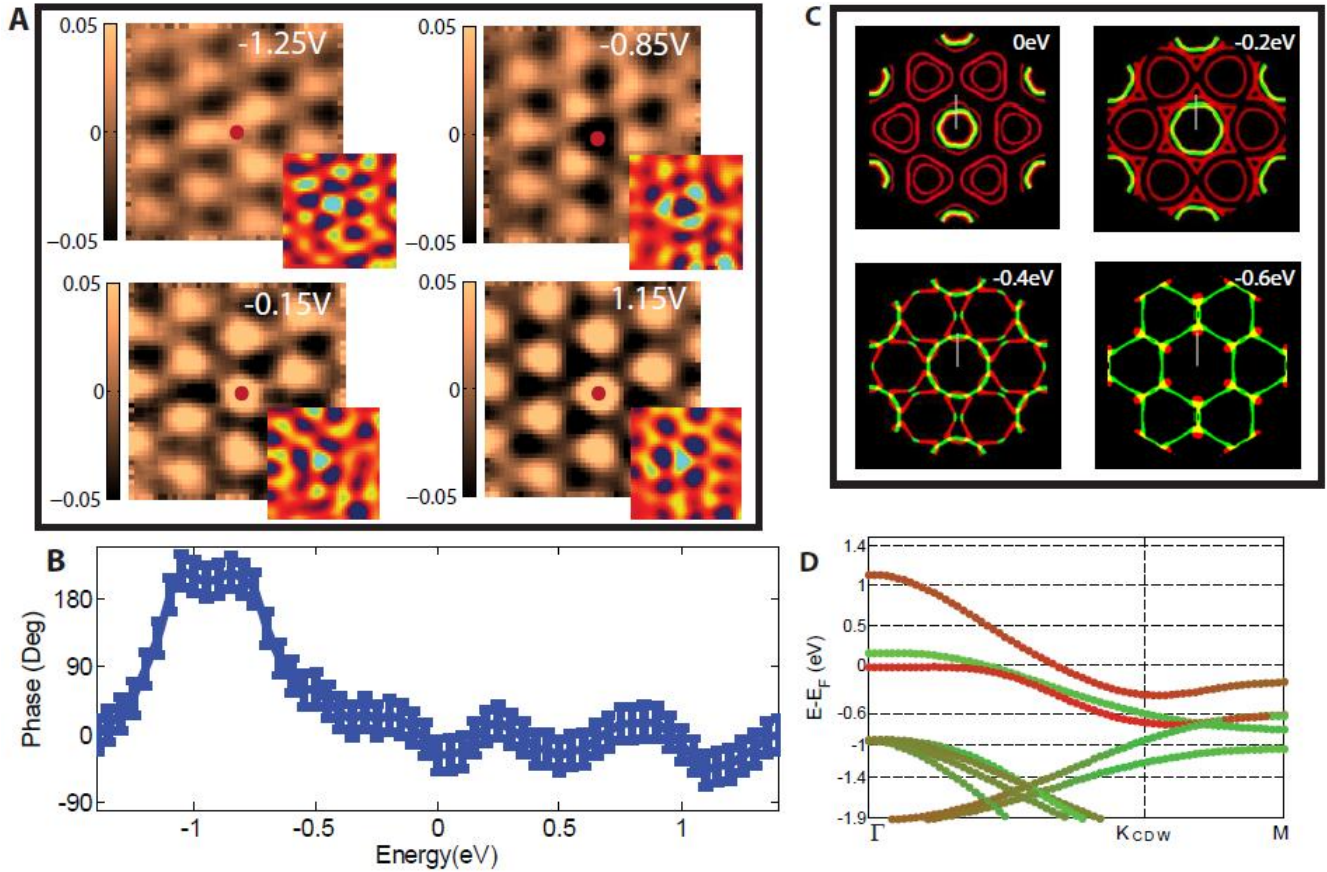

Fig. 4 (a) Cross correlation images between CDW band-pass filtered dI/dV maps taken at different energies at T=57°K, and filtered dI/dV map taken at V=1.4 V. The origin is indicated by a red dot. The CDW pattern changes from in phase (bright peak at the origin) to out of phase (origin is dark) between -0.65 and -1 V. The insets show the filtered dI/dV data in the vicinity of one defect that also illustrates this effect. (b) Extracted phase of the CDW at different energies relative to the phase at 1.4 V. Error bars given by the resolution limit of the image in pixels/Deg. (c) Calculated DOS (Nb-like orbitals in red, Se-like orbitals in green) at different energies. At energies close to -0.6 eV, significant intensity is present at the CDW wavevectors (large nesting) whereas this is not true at other energies shown. (d) Calculated band structure projection onto Nb(red) and Se(green) orbitals along the CDW direction showing that conditions for a gap opening at the CDW Bragg point are in the vicinity of -0.6 eV.

# SUPPLEMENTARY MATERIAL

# Visualizing the Charge Density Wave Transition in 2H-NbSe$_2$ in Real Space

S.P.Chockalingam[1,*], C.J.Arguello[1,*], E.P.Rosenthal[1], L. Zhao[1], C.Gutiérrez[1], J.H.Kang[1], W.C.Chung[1], R.M.Fernandes[1], S.Jia[2], A.J.Millis[1], R.J.Cava[2], A.N.Pasupathy[1,$]

[1]Department of Physics, Columbia University, New York NY 10027
[2]Department of Chemistry, Princeton University, Princeton NJ 08540
* These authors contributed equally to this work
$ Contact: apn2108@columbia.edu

## CDW autocorrelation images

In order to calculate the 2D autocorrelation images of the CDW (Fig. 2**c-f**), we start with the topographic image at each temperature as shown in Fig. S1**a**. The absolute value of the Fourier transform of this image is shown in Fig. S1**b**. A band pass filter is applied to the Fourier transform as shown in Fig. S1**c**, preserving only wave vectors near the CDW wave vector. The bandwidth of the filter is several times the standard deviation of the CDW peak. The filtered Fourier image is then transformed back to real space, giving the image shown in Fig. S1**d**. Finally, the 2D autocorrelation of this image is taken and symmetrized along the high-symmetry directions to produce the final figures shown in Figs. 2**c-f**.

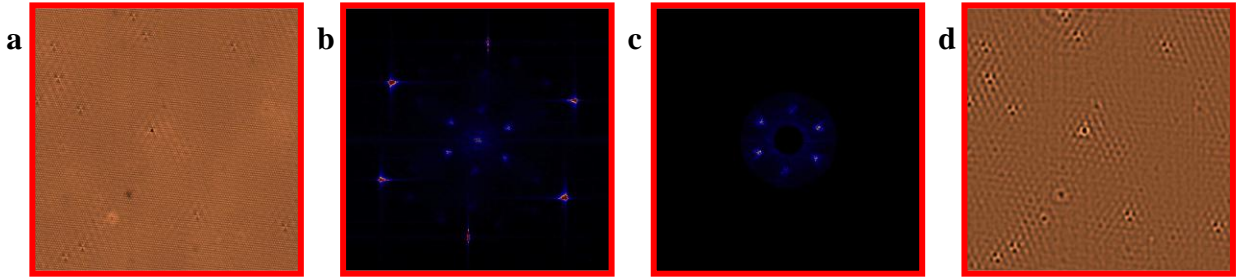

*Fig. S1.* **Autocorrelation procedure.** *a* Original topography. *b* Fourier transform of *a*. *c* Filtered fourier transform after removing the atomic and low frequencies. *d* real space image after the inverse Fourier transform of the filtered image.

## Temperature dependence of the Correlation length

To extract the correlation length from the autocorrelation images shown in Figs. 2 **c-f**, we first take a line cut along the x-axis through the origin of the image. Each of these line cuts (also shown in Figs. 2 **c-f**) can be fit to a function of the form

$$A(x) = A_0 \exp\left(-\frac{x}{\xi}\right) \cos(k_{CDW} x) + B_0$$

Here $B_0$ is the average of the random noise in the autocorrelation image ($B_0 \ll 1$) and $A_0 = 1 - B_0$.

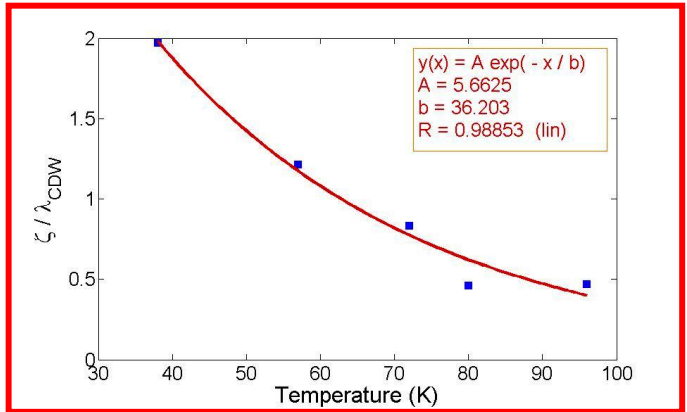

*Fig. S2: Extracted CDW correlation length as a function of temperature*

The extracted correlation length $\xi$ is plotted as a function of temperature in Fig. S2.

### DFT calculations of the band structure for NbSe$_2$

First principles calculations of the band structure of 2H-NbSe$_2$ were performed using DFT within the Local Density Approximation (LDA) with the Perdew-Zunger fit for the exchange correlation functional. All our calculations were carried out by using the Quantum Espresso package[1]. For Nb we used an ultrasoft Vanderbilt pseudopotential (Nb.pz-sp-van.UPF) with 13 valence electrons. For Se we used a norm-conserving potential (Se.pz-bhs.UPF) with 6 valence electrons. The cutoff energy for the basis set was chosen to be 30 Ry, and for the charge density of 180 Ry. The values for valid cutoff energies were determined by testing the energy convergence relevant to the system within a reasonable tolerance for our purposes (differences of less than 10meV with a 5X increase in the cutoff values). The unit cell used for the DFT calculations is shown in Fig. S3. It is composed of two equivalent but shifted Se-Nb-Se sub-units. The structure was allowed to relax until the forces were less than $10^{-3}$ Ry/a.u. The relaxed unit cell has a single layer Se-Se distance of 3.43 Å and a distance of 3.36 Å between top and bottom Se layers of the same sub-unit of the cell. The shortest Nb-Se distance is 2.60 Å. The minimum Nb-Nb distance cell is 6.26 Å. Our DFT results compare well with previous calculations[2] and with recent ARPES measurements of the Fermi surface[3]. The JDOS shown in Figs. 3G-J are autocorrelation images of the LDOS obtained from the DFT calculations.

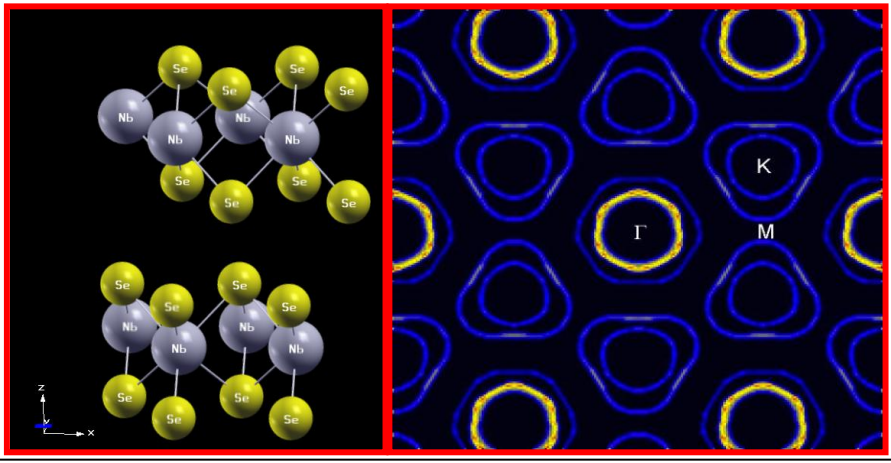

*Fig. S3: (left) Crystal structure for 2H-NbSe$_2$. The unit cell has two equivalent but shifted sub-layers. (right) Fermi surface from DFT calculations.*

# Phase Shift and Topography Images

The observed phase change of the CDW in the spectroscopy could originate from the constant current normalization condition involved in the STS measurements. In order to discard a possible unphysical origin of the phase change ($\phi_2-\phi_1$), we took several topographic measurements of the same area around defects for a wide range of bias (-1V to 1V).

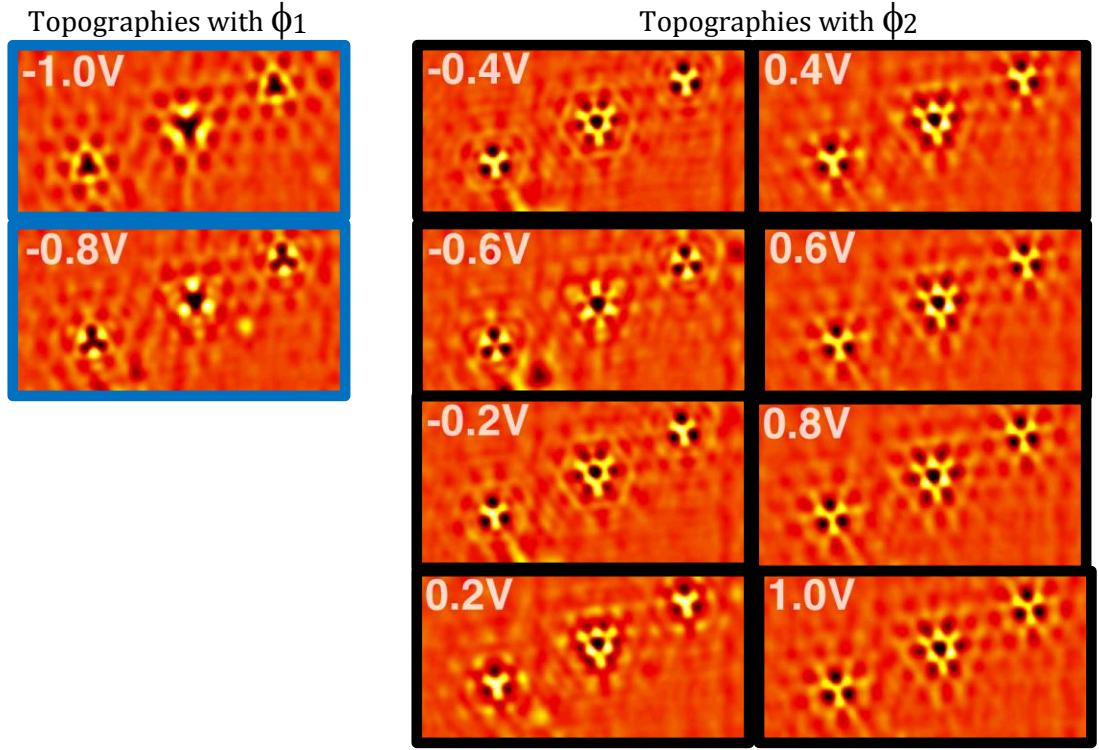

*Fig. S4. **Phase change on Topography images.** Topography images taken from -1V to 1V. Atoms have been filtered to enhance CDW. Images have been grouped by CDW phase in the left and right columns.*

From these images is clear that the phase remains constant for biases greater than -0.6V. The phase change is localized somewhere between -0.6 and -0.8eV.

## Tight Binding Model

To understand the changes in the phase of the CDW observed in our STS measurements, we developed a minimal one-dimensional tight-binding model that captures the essential features of the system. In the normal state, each unit cell contains two types of atoms, which we associate with Se and Nb (red and blue in Fig. S5, respectively). The onsite energies at each atom are denoted by $\Delta_1$ and $\Delta_2$, respectively; the nearest-neighbor hopping parameter (i.e. between two atoms of different types) is given by $t$, and the next-nearest neighbor hopping parameters (i.e. between two atoms of the same type) are denoted by $t'_1$ and $t'_2$.

In the CDW state, two consecutive atoms of the same type can have different onsite energies $\Delta_i - \delta_i$ and $\Delta_i + \delta_i$, while two consecutive bonds between different atoms can have different hopping parameters $t + \tau$ and $t - \tau$ (see Fig. S5). In either case, the unit cell in the CDW state doubles its size. If $\delta \neq 0$ and $\tau = 0$, we have a site-centered CDW, whereas in the case of $\tau \neq 0$ and $\delta = 0$ we have a bond-centered CDW.

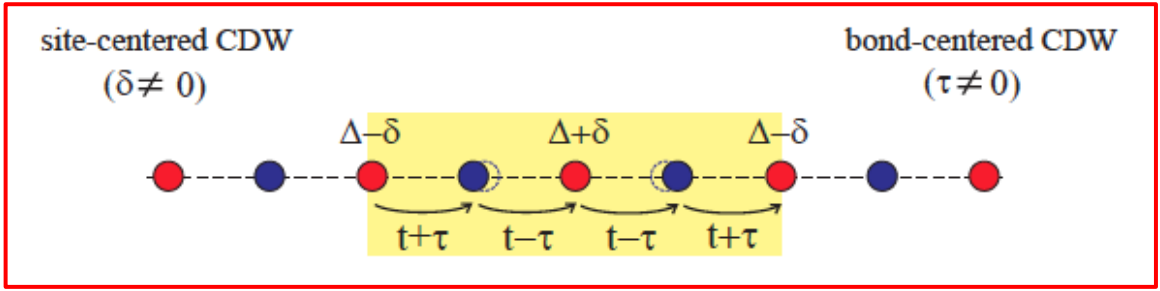

*Fig. S5.* **Unit cell used in the tight-binding model.** *The normal state parameters are the onsite energy ($\Delta$) and the nearest-neighbor hopping (t). The charge density wave parameters are the alternation in ``Se'' (red) site energy ($\delta$) and alternation in bond strength($\tau$).*

In the dI/dV measurements performed in the experiment, electrons tunnel into the Se atoms, which are the ones exposed after cleaving the crystal. The observed CDW contrast is the difference in the dI/dV signal between the peak and trough of the CDW in real space. In our model, this corresponds to the difference in the local density of states on adjacent Se atoms, and is given by:

$$\delta\left(\frac{dI}{dV}\right) = \sum_{k,\lambda}\left(\left|u_{k,\lambda}^{(Se)}\right|^2 - \left|u_{k+Q,\lambda}^{(Se)}\right|^2\right)\delta(V - \varepsilon_{k,\lambda}) \quad (S1)$$

where $u_{k,\lambda}^{(Se)}$ is the Se-atom component of the eigenfunction associated with the energy state $\varepsilon_{k,\lambda}$. A sign change in this quantity corresponds to a 180° change in the phase of the CDW, as seen in the STS maps. Here, $k$ is the momentum, $\lambda$ is the band-index, and $Q$ is the modulation vector of the charge-density wave.

In the normal state, $\left|u_{k,\lambda}^{(Se)}\right|^2 = \left|u_{k+Q,\lambda}^{(Se)}\right|^2$ and the contrast is zero, as expected. To study how these eigenfunctions change in the CDW state, it is enough to focus at the Bragg points $\varepsilon_{k,\lambda} = \varepsilon_{k+Q,\lambda}$, where the CDW gap opens. For simplicity, hereafter we assume that only the variation of the red (``Se'') site energy $\delta_1 \neq 0$. Then, after defining $\Delta \equiv \Delta_1 - \Delta_2$, the Hamiltonian of the system at the Bragg points is given by:

$$\mathcal{H}_{\text{Bragg}} = \Psi^{\dagger}_{k_{\text{Bragg}}} \begin{pmatrix} 2(\Delta+\delta) & 0 & -(t-\tau) & t-\tau \\ 0 & 2(\Delta-\delta) & -(t+\tau) & -(t+\tau) \\ -(t-\tau) & -(t+\tau) & 0 & 0 \\ t-\tau & -(t+\tau) & 0 & 0 \end{pmatrix} \Psi_{k_{\text{Bragg}}} \quad (S2)$$

where we defined the creation operators:

$$\Psi_k^{\dagger} = (c^{\dagger}_{k,Se} \quad c^{\dagger}_{k+Q,Se} \quad c^{\dagger}_{k,Nb} \quad c^{\dagger}_{k+Q,Nb}) \quad (S3)$$

We study separately the two possible CDW scenarios, i.e. site-CSW ($\delta \neq 0$ and $\tau = 0$) and bond-CDW ($\delta = 0$ and $\tau \neq 0$). To calculate the eigenfunctions, we use standard degenerate first-order perturbation theory with the unperturbed Hamiltonian $H_0 = H_{Bragg}(\delta = \tau = 0)$, and the perturbations $H_1 = H_{Bragg}(\delta \ll \Delta, \tau = 0) - H_0$ for site-CDW and $H_1 = H_{Bragg}(\delta = 0, \tau \ll t) - H_0$ for bond-CDW.

The diagonalization of $H_0$ yields two sets of doubly-degenerate energy eigenstates, corresponding to the upper and lower bands:

$$E_U \equiv \Delta + \sqrt{2t^2 + \Delta^2} > 0 \; ; \; E_L \equiv \Delta - \sqrt{2t^2 + \Delta^2} < 0 \quad (S4)$$

each with eigenvectors:

$$u_{i,1} = (E_i \quad 0 \quad -1 \quad 1) \; ; \; u_{i,2} = (0 \quad E_i \quad 1 \quad 1), \; i = L, U \quad (S5)$$

This degeneracy is lifted by either site or bond perturbations, which are both diagonal on the basis set of Eq. (S5). However, each case yields different results for the sign-change of the contrast across the two CDW gaps.

For concreteness, we consider the case $\delta>0$ and $\tau>0$. For site-CDW, the eigenvalue associated with the eigenstate $u_{i,1}$ is always greater than the one associated with $u_{i,2}$:

$$E_{i,(1,2)} = E_i \pm \delta \frac{E_i^2}{|E_i|\sqrt{2t^2 + \Delta^2}} \quad (S6)$$

since the first-order correction is proportional to the square of the unperturbed energy values. This implies that, for both upper and lower bands, the highest energy state has $|u_{k,\lambda}^{(Se)}|^2 > |u_{k+Q,\lambda}^{(Se)}|^2$, whereas the lowest energy state has $|u_{k,\lambda}^{(Se)}|^2 < |u_{k+Q,\lambda}^{(Se)}|^2$. Therefore, the contrast $\delta\left(\frac{dI}{dV}\right)$ changes its sign from negative to positive when the CDW gap is crossed from lower to higher energies. As a result, in the site centered case the contrast has to change signs and odd number of times between the two energies corresponding to the two gaps.

On the other hand, for bond-CDW, the first-order correction to each eigenvalue is linearly proportional to its unperturbed values:

$$E_{i,(1,2)} = E_i \mp \tau \frac{2E_i t}{|E_i|\sqrt{2t^2 + \Delta^2}} \quad (S7)$$

Thus, for the lower band, since $E_L < 0$, the eigenvalue associated with $u_{L,1}$ is greater than the eigenvalue associated with $u_{L,2}$, implying that the contrast changes its sign from negative to positive when the gap is crossed from lower to higher energies. The opposite is true for the upper band, and the contrast changes its sign from positive to negative when coming from lower energies. Consequently, in the bond centered case the contrast must either not change its sign or change it an even number of times between the energies where the two CDW gaps open. The results of the calculation are summarized in Fig. S6.

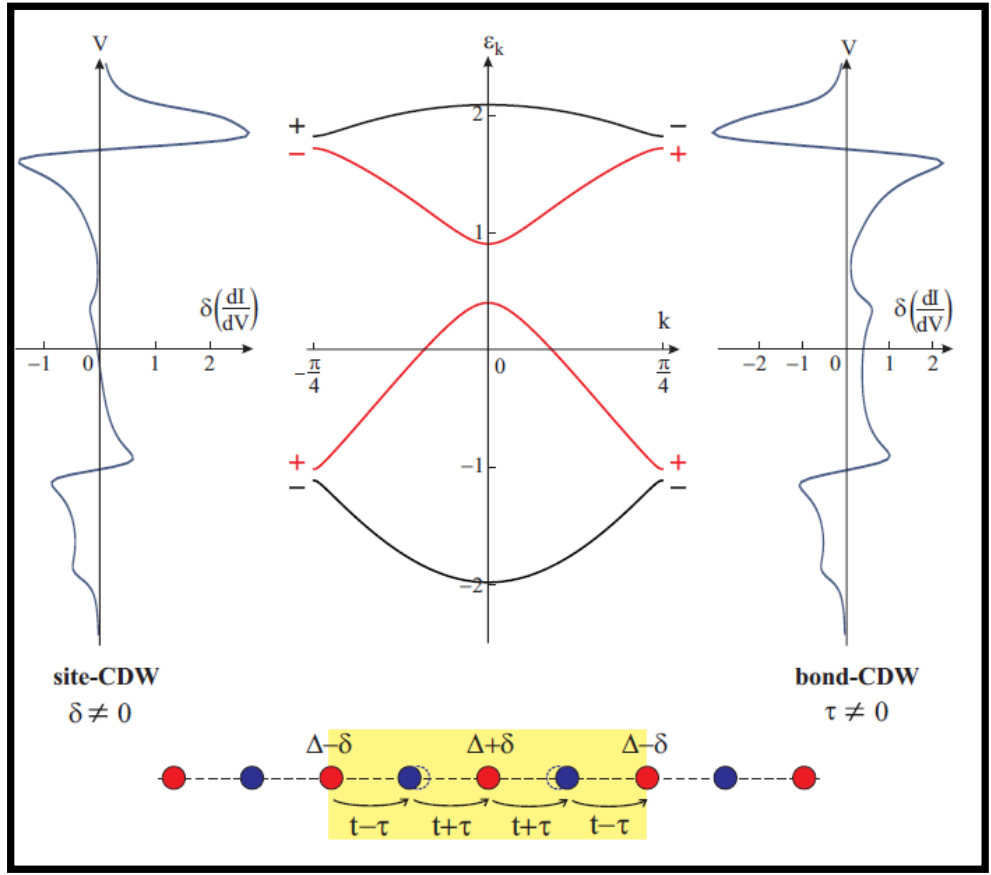

*Fig. S6. **Summary of the 1D tight-binding model results.** The two different types of CDW lead two different phase behaviors. The phase change seen in the STS maps corresponds to a site-CDW (odd number of phase changes). A gap at $\vec{K}_{CDW}$ for $E \simeq -0.7\,eV$ is to be expected from our experimental results.*

Notice that, in both site-CDW and bond-CDW cases, a large contrast is associated with the existence of a CDW gap at the energy where the contrast reversal occurs, and the magnitude of the contrast reversal is proportional to the size of the CDW gap.

# Comparison of Spectrum Curves

The phase change observed offers an indirect proof of the existence of a CDW at energies much below the Fermi Energy (-0.7eV). A more direct proof would come from the STS spectrum curves. More specifically, in a measurement with enough energy resolution, a comparison between curves located in regions with strong CDW order regions with no static order ought to show different features at those energies where a CDW gap opening is suspected.

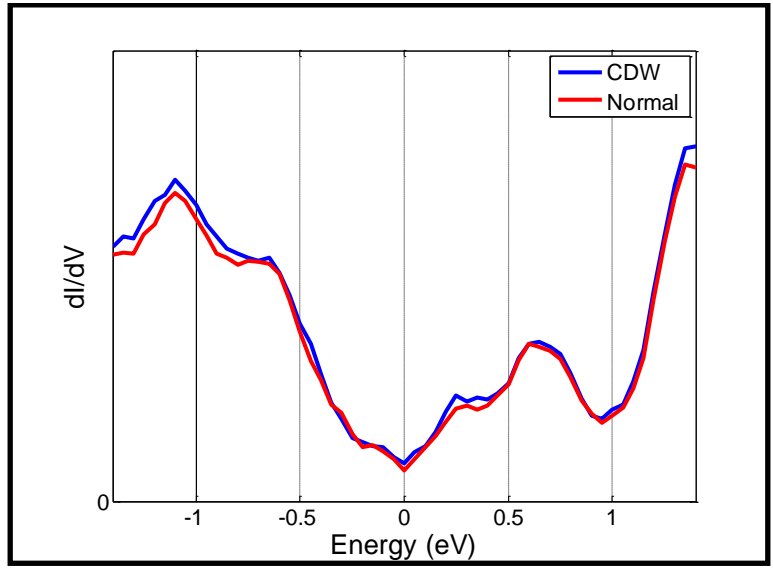

*Fig. S7. **Spectra comparison for CDW and normal states.** The CDW spectrum was chosen for a point in space with strong static order.*

A comparison between the spectrum of the normal and the CDW states is shown in Fig. S7. No obvious differences are present in the data. This is expected if the CDW gap opening is only for small regions in K-space. The large negative energies for which the gap is expected also make difficult to distinguish features associated with the existence of such gap.

## DFT Fat bands and nesting

To calculate the 2D "fat-bands" for $NbSe_2$, we obtained the projections of each DFT orbitals on Nb and Se eigenvectors, and calculated the Nb/Se character by adding the square of the corresponding projections. We then assigned a number ranging from 0(pure Nb) to 1(pure Se) in proportion to the projections obtained. An energy window of 15 meV was used to define the surfaces. Finally, to generate Figs. 5B-F the Nb/Se character of each eigenvalue was converted to a color scale, with pure Nb in red and pure Se in green.

As described in the main text, appropriate conditions for nesting at the observed CDW wave vector are found around -0.6 eV in our calculations. Shown in Fig. S8 are the fat bands for energies between -0.1 and +0.4 eV, with the CDW wave vector indicated by a white arrow. The nesting condition is totally absent for the positive range of energies.

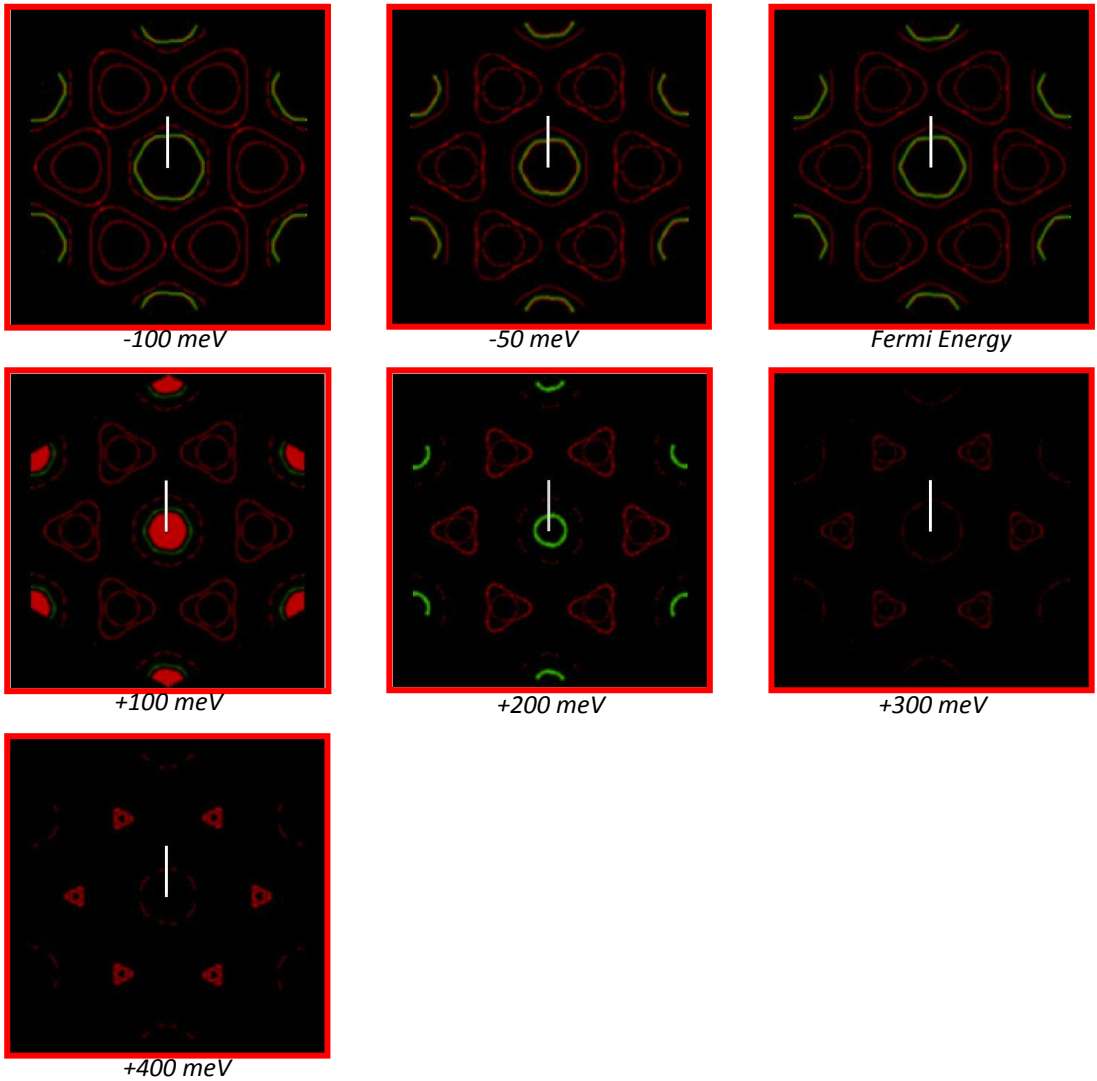

*Fig. S8.* **Fat bands calculated from -100meV to 400meV (Nb in red, Se in green).** *One of the CDW vectors is shown in white.*